\documentclass[english]{article}
\usepackage[T1]{fontenc}
\usepackage[latin9]{inputenc}
\usepackage{color}
\usepackage{units}
\usepackage{textcomp}
\usepackage{amstext}

\makeatletter
\date{}

\makeatother

\usepackage{babel}
\begin{document}

\title{On the Quantum Chromodynamics of a Massive Vector Field in the Adjoint
Representation}

\author{Alfonso R. Zerwekh%
\thanks{alfonso.zerwekh@usm.cl%
}\\
\textit{\normalsize Departamento de Física and }\\
\textit{\normalsize Centro Científico-Tecnológico de Valparaíso}\\
\textit{\normalsize Universidad Técnica Federico Santa María}\\
\textit{\normalsize Casilla 110-V, Valparaíso, Chile}}
\maketitle
\begin{abstract}
In this paper, we explore the possibility of constructing the quantum
chromodynamics of a massive color-octet vector field without introducing
higher structures like extended gauge symmetries, extra dimensions
or scalar fields. We show that gauge invariance is not enough to constraint
the couplings. Nevertheless the requirement of unitarity fixes the
values of the coupling constants, which otherwise would be arbitrary.
Additionally, it opens a new discrete symmetry which makes the coloron
stable and avoid its resonant production at a collider. On the other
hand, a judicious definition of the gauge fixing terms modifies the
propagator of the massive field making it well-behaved in the ultra-violet
limit. The relation between our model and the more general approach
based on extended gauge symmetries is also discussed.
\end{abstract}

\section{Introduction}

Many extensions of the Standard Model, such as non-minimal Technicolor
\cite{EHLQ,Lane-Ramana,Zerwekh-2004,Zerwekh-2007,Zerwekh-Rosenfeld},
Extra-dimensions \cite{KK-gluon} , Top-color \cite{Coloron1,Coloron2}
and Chiral-color \cite{Chiral-color1,Chiral-Color2,Chiral-Color3,Chiral-Color4}
, predict the existence of massive color-octet spin-1 particles which
we will collectively call ``colorons''. In principle, it is expected
that, if a coloron exists in the appropriated mass range, it should
be copiously produced at hadron colliders such as the Tevatron or
the LHC \cite{Zerwekh-2007,Dobrescu}. Indeed, some renewed interest
on this kind of particles has arisen \cite{Axigluon1,Axigluon2} because
some sort of color-octet spin-1 resonance may be the origin of the
large $t\bar{t}$ forward-backward asymmetry measured by CDF \cite{CDF1,CDF2}
and D0 \cite{D0}.

From the phenomenological point of view, it is convenient, given the
large variety of models predicting colorons, to find an effective
model-independent description which can grasp their essential features.
This is the origin, for example, of the deconstruction idea \cite{Deconstruction1,Deconstruction2}:
the initial intension was to describe the Kaluza-Klein excitations
of the gluon independently of the details of the underlying extra-dimensional
theory. A similar problem, but in the context of non-minimal Technicolor,
motivated another effective description \cite{Zerwekh-Rosenfeld,Zerwekh-2007}
based on the observed low-energy symmetries. This phenomenological
analysis produced two important results: it was shown that the $s$-channel
production of a single coloron is plagued by theoretical uncertainties
\cite{Zerwekh-2007} and it was argued that the coloron pair production
would be almost model-independent \cite{Zerwekh-2007,Dobrescu}, being
determined exclusively by QCD gauge invariance.

Interestingly, when one tries to make the quantum chromodynamics of
a coloron, one finds that there are at least two different, although
gauge equivalent, formulations \cite{Zerwekh-Rosenfeld,VGG}. Both
of them lead to theories with a bad ultraviolet behavior due to the
presence of the massive spin-1 field. This is a quite frustrating
situation because we know how to construct consistent and renormalizable
quantum field theories with scalars or fermions (massive or not) as
matter fields but things seem to be very different when a massive
spin-1 field is considered.

In this work, we revisit the construction of a gauge theory for the
coloron and we examine the possibility that such a theory be consistent
with renormalizability and unitarity without introducing neither scalar
fields nor higher structures such as extra-dimensions or extended
gauge symmetries. It is important to emphasize that our aim is not
to present an alternative to the well established method of introducing
colorons through the breaking down of a enlarged gauge sector, but
rather to investigate what are the minimal requirements for a consistent
coloron model only in the framework of the observed QCD gauge symmetry.
For this purpose, we organized this paper in the following way. In
section \ref{sec:A-Gauge-Theory}, we describe the construction of
a general classical gauge theory with a massive spin-one field in
the adjoint representation. In section \ref{sec:BRST}, we move to
the quantum version of the theory, paying special attention to the
gauge fixing and the ghost terms, and implementing the BRST symmetry.
Section \ref{sec:Unitarity} is devoted to study the constraints imposed
by requiring that perturbative unitarity of the S-matrix holds at
tree-level. Finally, we summarize our conclusions in section \ref{sec:Summary-and-Conclusions}.

\section{A Gauge Theory for a Massive Vector Field \label{sec:A-Gauge-Theory}}

\subsection{Global Symmetry}

Usually, the starting point for studying the physical properties of
a massive spin-one field is the Proca Lagrangian. So, let's consider
a generalization of the Proca theory for a non-Abelian \textcolor{black}{\emph{global}}
continuous symmetry:

\begin{equation}
\mathcal{L}=-\frac{1}{4}F_{\mu\nu}^{a}F^{a\mu\nu}+\frac{1}{2}M^{2}V_{\mu}^{a}V^{a\mu}-V_{\mu}^{a}J^{a\mu}(+\mathcal{L}_{int})\label{eq:Proca}
\end{equation}
 where $F_{\mu\nu}^{a}=\partial_{\mu}V_{\nu}^{a}-\partial_{\nu}V_{\text{\ensuremath{\mu}}}^{a}$
and $V_{\mu}^{a}$ transforms homogeneously under the global symmetry
($V_{\mu}\rightarrow U^{\dagger}V_{\mu}U$). We have included an external
source ($J^{a\mu}$) which is supposed to be a conserved current.
Additionally, all the other invariant terms that can be constructed
with $V_{\mu}$ and $\partial_{\mu}V_{\nu}$ can be eventually included
in $\mathcal{L}_{int}$, but they will not be relevant for the present
analysis and, for simplicity, will not be taken explicitly into account.
As it is well known, the field equation obtained from Lagrangian (\ref{eq:Proca})
can be written as:

\begin{equation}
\partial_{\rho}F^{a\rho\nu}+M^{2}V^{a\nu}=J^{a\nu}
\end{equation}
 and the anti-symmetry of $F_{\mu\nu}$ automatically implies the
Lorenz condition:

\begin{equation}
\partial_{\mu}V^{a\mu}=0\label{eq:LonrenzCondition}
\end{equation}
Let us recall, for completeness, that equation (\ref{eq:LonrenzCondition})
eliminates one degree of freedom from $V_{\mu}$. Naturally, the remaining
degrees of freedom correspond to the three polarization states of
a massive spin-one particle. 

Unfortunately, when a quantum theory is constructed from Lagrangian
(\ref{eq:Proca}), it leads to the following propagator:
\begin{equation}
\Delta_{\mu\nu}=\frac{-i}{q^{2}-M^{2}}\left(g_{\mu\nu}-\frac{q_{\mu}q_{\nu}}{M^{2}}\right)\label{eq:ProcaPorpagator}
\end{equation}
 which spoils the renormalizability of the theory due to the bad ultraviolet
behavior of the last term. Interestingly, the form of this propagator,
and its ultraviolet divergence, can be traced to the anti-symmetric
structure of $F_{\mu\nu}$. So, we can raise the reasonable question
of whether an anti-symmetric $F_{\mu\nu}$ is essential to our theory,
or not. In other words, is there a fundamental principle that compels
us to include $\partial_{\mu}V_{\nu}\partial^{\mu}V^{v}$ and $\partial_{\mu}V_{\nu}\partial^{\nu}V^{\mu}$
in the Lagrangian with the same weight ? Certainly, the answer is
negative. In gauge theory, the anti-symmetry of $F_{\mu\nu}$ is dictated
by the gauge principle since its necessary to cancel the inhomogeneous
part of the transformation of the gauge filed, but this is not the
case here because in our construction $V_{\mu}$ transforms homogeneously.
Consequently, it is possible to write down a more general Lagrangian:
\begin{equation}
\mathcal{L}=-\frac{1}{2}\partial_{\mu}V_{\nu}^{a}\partial^{\mu}V^{av}+\frac{\left(1+a\right)}{2}\partial_{\mu}V_{\nu}^{a}\partial^{\nu}V^{a\mu}+\frac{1}{2}M^{2}V_{\mu}^{a}V^{a\mu}-V_{\mu}^{a}J^{a\mu}(+\mathcal{L}_{int})\label{eq:StueckelberLagrangian}
\end{equation}
 Of course, the Proca Lagrangian is recovered for $a=0$. This time,
the field equation is:

\begin{equation}
\partial^{2}V_{\mu}-a\partial_{\mu}\partial^{\sigma}V_{\sigma}+M^{2}V_{\mu}=J_{\mu}\label{eq:StueckelberEquation}
\end{equation}
where we have dropped the group index. Differentiating (\ref{eq:StueckelberEquation}),
it follows that a generalized Lorenz condition is satisfied: 
\begin{equation}
\partial_{\mu}V^{\mu}=f(x)\label{eq:GeneralLorenzCondition}
\end{equation}
where $f(x)$ is a solution of the following equation:

\begin{equation}
\left[a\partial^{2}-M^{2}\right]f(x)=0\label{eq:ConsistencyCondition}
\end{equation}
Of course, $f(x)=0$ is a solution of (\ref{eq:ConsistencyCondition})
and the usual Lorenz condition can be used.

A more important consequence, however, is the fact that the propagator
obtained from (\ref{eq:StueckelberLagrangian}) can be written in
following way:
\begin{equation}
\Delta_{\mu\nu}=\frac{-i}{q^{2}-M^{2}}\left(g_{\mu\nu}-\frac{\left(1+a\right)q_{\mu}q_{\nu}}{aq^{2}-M^{2}}\right)\label{eq:StueckelbergPropagator}
\end{equation}

Notice that this new propagator has the same form of the propagator
of a massive gauge boson in the context of spontaneously broken gauge
symmetries. This modified propagator behaves adequately in the ultraviolet
limit. 

In the Abelian case, the theoretical construction presented so far
is similar to the one resulting from the Stueckelberg theory when
the compensating scalar field is gauged away. Indeed, equations (\ref{eq:StueckelberEquation})
and (\ref{eq:ConsistencyCondition}) are formally equal to those obtained
from the Stueckelberg Lagrangian \cite{Itzykson-Zuber}. It is worth
to recall that the Stueckelberg formalism makes the theory of a massive
photon renormalizable. For this reason, Lagrangian (\ref{eq:StueckelberLagrangian})
seems to be a good starting point in the attempt of making a consistent
theory for the coloron.

\subsection{Local Symmetry}

Evidently, the most direct way to turn the previous construction into
a local gauge theory is to take Lagrangian (\ref{eq:StueckelberLagrangian}),
replace partial derivatives by covariant ones, include a Yang-Mills
term for the gluon, include in $\mathcal{L}_{int}$ all the gauge
invariant and renormalizable terms we can form with $V_{\mu}$, $D_{\mu}V_{\nu}$
and $G_{\mu\nu}$ (field-strength of gluons) with arbitrary coefficients.
In principle, this Lagrangian should contain a kinetic mixing term,
nevertheless, it can be removed by a simple redefinition of the fields
(see, for instance, \cite{VGG}). This standard procedure leads us
to the following Lagrangian:

\begin{eqnarray}
\mathcal{L} & = & -\frac{1}{2}Tr\left\{ G_{\mu\nu}G^{\mu\nu}\right\} -Tr\left\{ D_{\mu}V_{\nu}D^{\mu}V^{\nu}\right\} +\left(1+a\right)Tr\left\{ D_{\mu}V_{\nu}D^{\nu}V^{\mu}\right\} \nonumber \\
 &  & +a_{11}Tr\left\{ D_{\mu}V_{\nu}V^{\mu}V^{\nu}\right\} +a_{12}Tr\left\{ D_{\mu}V_{\nu}V^{\nu}V^{\mu}\right\} \label{eq:LagPhysFields}\\
 &  & +a_{21}Tr\left\{ V_{\mu}V_{\nu}V^{\mu}V^{\nu}\right\} +a_{22}Tr\left\{ V_{\mu}V_{\nu}V^{\nu}V^{\mu}\right\} \nonumber \\
 &  & +a_{3}Tr\left\{ G_{\mu\nu}\left[V^{\mu},V^{\nu}\right]\right\} +M^{2}Tr\{V_{\nu}V^{\nu}\}\nonumber 
\end{eqnarray}

We recall that this Lagrangian is invariant under the local gauge
transformations:

\begin{eqnarray*}
G_{\mu} & \rightarrow & UG_{\mu}U^{-1}-\frac{1}{g}\left(\partial_{\mu}U\right)U^{-1}\\
V_{\mu} & \rightarrow & UV_{\mu}U^{-1}
\end{eqnarray*}

Notice that the third term in (\ref{eq:LagPhysFields}) includes the
implementation of the Stueckelberg trick and, in the quantum version
of the theory developed so far, the propagator of the coloron would
be the one shown in (\ref{eq:StueckelbergPropagator}). Nevertheless,
the same term modifies the $GVV$ and the $GGVV$ vertex (the last
one is also modified by the term proportional to $a_{3}$). This fact
is important because it means that it is not guaranteed that, in the
general case, a massive color-octet spin-one particle interacts with
gluons with a typical QCD strength as it is commonly believed. 

At this point it is important to notice that this Lagrangian is written
in a very specific basis, which we call the physical basis, where
the mass matrix is diagonal and the massive field transforms homogeneously
under gauge transformations. Nevertheless, it will be convenient for
our purposes to write Lagrangian (\ref{eq:LagPhysFields}) in terms
of a general basis. For this reason, we define new fields $A_{1\mu}^{a}$
and $A_{2\mu}^{a}$, by rotating $G$ and $V$, in such a way that

\begin{eqnarray}
G_{\mu} & = & \frac{g_{2}}{\sqrt{g_{1}^{2}+g_{2}^{2}}}A_{1\mu}+\frac{g_{1}}{\sqrt{g_{1}^{2}+g_{2}^{2}}}A_{2\mu}\\
V_{\mu} & = & \frac{g_{1}}{\sqrt{g_{1}^{2}+g_{2}^{2}}}A_{1\mu}-\frac{g_{2}}{\sqrt{g_{1}^{2}+g_{2}^{2}}}A_{2\mu}
\end{eqnarray}

In these expressions $g_{1}$ and $g_{2}$ are constant that satisfy
the constrain

\begin{equation}
g\equiv\frac{g_{1}g_{2}}{\sqrt{g_{1}^{2}+g_{2}^{2}}}
\end{equation}

where $g$ is the usual QCD coupling constant.

In terms of the new basis, the Lagrangian (\ref{eq:LagPhysFields})
can be re-organized as:

\begin{eqnarray}
\mathcal{L} & = & -\frac{1}{2}Tr\left[F_{1\mu\nu}F_{1}^{\mu\nu}\right]-\frac{1}{2}Tr\left[F_{2\mu\nu}F_{2}^{\mu\nu}\right]\nonumber \\
 &  & +\frac{M^{2}}{g_{1}^{2}+g_{2}^{2}}Tr\left[\left(g_{1}A_{1\mu}-g_{2}A_{2\mu}\right)^{2}\right]\label{eq:L2CYM1}\\
 &  & +\frac{a}{g_{1}^{2}+g_{2}^{2}}Tr\left[\left(g_{1}D_{\mu}A_{1\nu}-g_{2}D_{\mu}A_{2\nu}\right)\left(g_{1}D^{\nu}A_{1}^{\mu}-g_{2}D^{\nu}A_{2}^{\mu}\right)\right]\nonumber \\
 &  & +\mathcal{L}_{NM}\nonumber 
\end{eqnarray}

where

\[
F_{j\mu\nu}=\partial_{\mu}A_{j\nu}-\partial_{\nu}A_{j\mu}-ig\left[A_{j\mu},A_{j\nu}\right]
\]

and

\[
D_{\mu}=\partial_{\mu}-i\frac{g^{2}}{g_{1}}\left[A_{1\mu},\:\right]-i\frac{g^{2}}{g_{2}}\left[A_{2\mu},\:\right]
\]

$\mathcal{L}_{NM}$ represents all the interaction terms depending
on arbitrary constants.

The gauge transformation of the new fields can be easily obtained
form the transformation laws of the gluon and the massive field $V_{\mu}$.
The resulting transformation law is:

\begin{equation}
A_{i\mu}\rightarrow UA_{i\mu}U^{-1}-\frac{1}{g_{i}}\left(\partial_{\mu}U\right)U^{-1}\quad\left(i=1,2\right)\label{eq:Atranformation}
\end{equation}

This means that the fields $A_{1\mu}^{a}$ and $A_{2\mu}^{a}$ transform
like connections. Of course, Lagrangian (\ref{eq:L2CYM1}) is invariant
by construction under this gauge transformation. The important point
is that now the coloron Lagrangian is written in such a way that it
formally seems a gauge theory with two connections. In the next section
we will use this fact to write the quantum version of the theory and
implement the BRST symmetry by applying twice the Fadeev-Popov prescription.

\section{Quantum Theory: BRST Symmetry \label{sec:BRST}}

Strictly speaking, what we have done so far is to develop a classical
theory. If we want to quantize the theory using the path integral
method, it is necessary to add gauge fixing terms and ghost fields
as dictated by the Fadeev-Popov procedure. Fortunately, this is an
easy task in the version of the model developed in the previous section.
A good starting point is Lagrangian (\ref{eq:L2CYM1}). Because we
have two gauge-like fields, all we need to do is to duplicate the
standard Fadeev-Popov prescription and add to (\ref{eq:L2CYM1}) the
following Lagrangian:
\begin{eqnarray}
\mathcal{L}_{GF} & = & \frac{1}{2}\xi_{1}B_{1}^{a}B_{1}^{a}-B_{1}^{a}\partial^{\mu}A_{1\mu}^{a}+\bar{c}_{1}^{a}\partial^{\mu}D_{1\mu}^{ab}c^{b}\nonumber \\
 &  & +\frac{1}{2}\xi_{2}B_{2}^{a}B_{2}^{a}-B_{2}^{a}\partial^{\mu}A_{2\mu}^{a}+\bar{c}_{2}^{a}\partial^{\mu}D_{2\mu}^{ab}c^{b}\label{eq:LGF1}
\end{eqnarray}
where, as usual, $B_{1}^{a}$ and $B_{2}^{a}$ are auxiliary fields,
$c$, $\bar{c}_{1}$and $\bar{c}_{2}$ are ghost and anti-ghost fields
and $\xi_{1}$ and $\xi_{2}$ are gauge parameters. Additionally,
we have used the notation $D_{j\mu}\equiv\partial_{\mu}-ig_{j}\left[A_{j\mu},\:\right]$.
In order to avoid the introduction of kinetic mixing terms in the
physical basis (because we expect that in the basis formed by $G_{\mu}$
and $V_{\mu}$ everything is diagonal), we chose $\xi_{1}=\xi_{2}=\xi$.
Of course, by construction, the whole Lagrangian (that is, (\ref{eq:L2CYM1})
+ (\ref{eq:LGF1}) ) is invariant under the BRST transformations:
\begin{eqnarray}
\delta_{B}A_{i\mu}^{a} & = & \frac{1}{g_{i}}D_{i\mu}^{ab}c^{b}\\
\delta_{B}c^{a} & = & -\frac{1}{2}f^{abc}c^{b}c^{c}\\
\delta_{B}\bar{c}_{i}^{a} & = & B_{i}^{a}\\
\delta_{B}B_{i}^{a} & = & 0
\end{eqnarray}

In the physical basis, Lagrangian (\ref{eq:LGF1}) takes the form:
\begin{eqnarray}
\mathcal{L}_{GF} & = & -\frac{1}{2\xi}\left(\partial^{\mu}G_{\mu}^{a}\right)^{2}-\frac{1}{2\xi}\left(\partial^{\mu}V_{\mu}^{a}\right)^{2}+\bar{c}^{a}\partial^{\mu}D_{\mu}^{ab}c^{b}\nonumber \\
 &  & +\alpha f^{abc}\left(\partial^{\mu}\bar{c}^{a}\right)V_{\mu}^{c}c^{b}+\beta f^{abc}\left(\partial^{\mu}\bar{\eta}^{a}\right)V_{\mu}^{c}c^{b}\label{eq:LGF2}
\end{eqnarray}
where we have already eliminated the auxiliary fields. Here $\alpha$
and $\beta$ are some combinations of the original coupling constants
$g_{1}$ and $g_{2}$ but their exact expressions are not important
for our purposes. On the other hand, $\bar{c}$ and $\bar{\eta}$
are related to the previous anti-ghosts by the following definitions:

\begin{eqnarray}
\bar{c} & \equiv & \bar{c}_{1}+\bar{c}_{2}\\
\bar{\eta} & \equiv & \bar{c}_{2}-\bar{c}_{1}
\end{eqnarray}

An important characteristic of Lagrangian (\ref{eq:LGF2}) is that
$\bar{\eta}$ doesn't have a kinetic term and hence its equation of
motion is only a constraint:
\begin{equation}
f^{abc}\partial^{\mu}\left(V_{\mu}^{c}c^{b}\right)=0\label{eq:etaconstrain}
\end{equation}
Interestingly, this is exactly the kind of constraint needed to implement
the Lorenz condition for a massive field transforming homogeneously
under the symmetry group. Putting (\ref{eq:etaconstrain}) back in
the Lagrangian, we find:
\begin{eqnarray}
\mathcal{L}_{GF} & = & -\frac{1}{2\xi}\left(\partial^{\mu}G_{\mu}^{a}\right)^{2}-\frac{1}{2\xi}\left(\partial^{\mu}V_{\mu}^{a}\right)^{2}+\bar{c}^{a}\partial^{\mu}D_{\mu}^{ab}c^{b}\label{eq:LGF3}
\end{eqnarray}
So finally, we have the correct gauge fixing and ghost terms for our
model. The most important consequence of this procedure is that the
second term of (\ref{eq:LGF3}) contributes to the propagator of $V_{\mu}$.
Indeed, we can write now the correct propagator for the gluon and
the coloron in the complete theory:
\begin{eqnarray}
\Delta_{G} & = & \frac{-i\delta^{ab}}{q^{2}}\left(g^{\mu\nu}+\left(\xi-1\right)\frac{q^{\mu}q^{\nu}}{q^{2}}\right)\\
\Delta_{V} & = & \frac{-i\delta^{ab}}{q^{2}-M^{2}}\left(g^{\mu\nu}+\left(\xi+\xi a-1\right)\frac{q^{\mu}q^{\nu}}{(1-\xi a)q^{2}-\xi M^{2}}\right)
\end{eqnarray}

\section{Unitarity \label{sec:Unitarity}}

\subsection{Unitarity Constraints }

Hitherto, we have constructed a general quantum gauge theory of the
coloron with operators of dimension four or less. It can be seen as
a good starting point for an effective theory and its main consequence
is that the coupling of the coloron to gluon may sensibly deviate
from the general expectation. However, we would like to recall that
the aim of this work is to explore the construction of a coloron theory
which can be well behaved in the ultraviolet limit and, eventually,
renormalizable. In this sense, it is necessary to compel the theory
to preserve the perturbative unitarity of the S-matrix. For this purpose,
we compute the amplitudes for the processes $V_{L}V_{L}\rightarrow V_{L}V_{L}$
and $GG\rightarrow V_{L}V_{L}$ (where $G$ is the gluon and $V_{L}$
is the longitudinally polarized coloron) at tree-level and we impose
the condition that the terms which are divergent in the ultraviolet
limit ($\nicefrac{s}{M^{2}}\rightarrow\infty$) vanish. In these calculations,
we use a simplified version of (\ref{eq:LagPhysFields}):
\begin{eqnarray}
\mathcal{L} & = & -\frac{1}{2}Tr\left\{ G_{\mu\nu}G^{\mu\nu}\right\} -Tr\left\{ D_{\mu}V_{\nu}D^{\mu}V^{\nu}\right\} +\left(1+a\right)Tr\left\{ D_{\mu}V_{\nu}D^{\nu}V^{\mu}\right\} \nonumber \\
 &  & +a_{1}Tr\left\{ \left(D_{\mu}V_{\nu}-D_{\nu}V_{\mu}\right)\left[V^{\mu},V^{\nu}\right]\right\} \nonumber \\
 &  & +a_{2}Tr\left\{ \left[V_{\mu},V_{\nu}\right]\left[V^{\mu},V^{\nu}\right]\right\} \\
 &  & +a_{3}Tr\left\{ G_{\mu\nu}\left[V^{\mu},V^{\nu}\right]\right\} +M^{2}Tr\{V_{\nu}V^{\nu}\}\nonumber \\
 &  & -\frac{1}{2\xi}\left(\partial^{\mu}G_{\mu}^{a}\right)^{2}-\frac{1}{2\xi}\left(\partial^{\mu}V_{\mu}^{a}\right)^{2}+\bar{c}^{a}\partial^{\mu}D_{\mu}^{ab}c^{b}\nonumber 
\end{eqnarray}
This simplification is well motivated, however, because the terms
in (\ref{eq:LagPhysFields}) containing three and four $V$ fields
would give origin in the amplitude to terms proportional to the $d^{abc}$
(the completely symmetric constants of the group) which should cancel
among themselves. The best way to assure this cancellation is organizing
the $V$ self-interactions in terms of the commutators. The divergent
part of the amplitude for the $V_{L}V_{L}\rightarrow V_{L}V_{L}$
scattering can be written as:
\begin{eqnarray}
\mathcal{M} & = & \frac{\left(a_{2}+g^{2}+a_{1}^{2}\right)\left(t^{2}-2tu-2u^{2}\right)\left(t+u\right)^{2}itu}{4s\left(M^{2}-s\right)\left(M^{2}-t\right)\left(M^{2}-u\right)M^{4}}f^{abe}f^{cde}\nonumber \\
 &  & +\frac{\left(a_{2}+g^{2}+a_{1}^{2}\right)\left(t^{2}+4tu+u^{2}\right)\left(t+u\right)^{2}itu}{4s\left(M^{2}-s\right)\left(M^{2}-t\right)\left(M^{2}-u\right)M^{4}}f^{ace}f^{bde}\nonumber \\
 &  & -\frac{\left(a_{2}+g^{2}\right)\left(t^{4}+11t^{3}u-23t^{2}u^{2}-28tu^{3}-2u^{4}\right)\left(t+u\right)i}{4s\left(M^{2}-s\right)\left(M^{2}-t\right)\left(M^{2}-u\right)M^{2}}f^{abe}f^{cde}\nonumber \\
 &  & -\frac{6i\left(t+u\right)^{2}ag^{2}t^{2}u}{4s\left(M^{2}-s\right)\left(M^{2}-t\right)\left(M^{2}-u\right)M^{2}}f^{abe}f^{cde}\nonumber \\
 &  & -\frac{\left(t^{4}+14t^{3}u-20t^{2}u^{2}-28tu^{3}-2u^{4}\right)a_{1}^{2}\left(t+u\right)i}{4s\left(M^{2}-s\right)\left(M^{2}-t\right)\left(M^{2}-u\right)M^{2}}f^{abe}f^{cde}\nonumber \\
 &  & -\frac{\left(t^{4}+14t^{3}u+40t^{2}u^{2}+14tu^{3}+u^{4}\right)a_{1}^{2}\left(t+u\right)i}{4s\left(M^{2}-s\right)\left(M^{2}-t\right)\left(M^{2}-u\right)M^{2}}f^{ace}f^{bde}\nonumber \\
 &  & +\frac{6i\left(t+u\right)^{3}ag^{2}tu}{4s\left(M^{2}-s\right)\left(M^{2}-t\right)\left(M^{2}-u\right)M^{2}}f^{ace}f^{bde}\nonumber \\
 &  & -\frac{\left(a_{2}+g^{2}\right)\left(t^{4}+17t^{3}u+46t^{2}u^{2}+17tu^{3}+u^{4}\right)\left(t+u\right)i}{4s\left(M^{2}-s\right)\left(M^{2}-t\right)\left(M^{2}-u\right)M^{2}}f^{ace}f^{bde}\nonumber \\
 &  & +{\mathcal{O}}\left(\frac{M^{2}}{s}\right)
\end{eqnarray}
It can be easily seen that, in order to cancel the problematic terms,
it is enough to satisfy the following conditions:

\begin{eqnarray}
a & = & 0\\
a_{1} & = & 0\\
a_{2} & = & -g^{2}\label{eq:a0}
\end{eqnarray}
Indeed, with this election of parameters, the amplitude for the $V_{L}V_{L}\rightarrow V_{L}V_{L}$
scattering completely vanishes. Surprisingly, the previous conditions
avoid the presence of the Stueckerberg term and forbid the coloron
triple vertex. 

In a similar way, imposing unitarity to the $GG\rightarrow V_{L}V_{L}$
scattering amplitude, we get an additional condition:

\begin{equation}
a_{3}=-g
\end{equation}
Consequently, taken into account the restrictions due to unitarity,
the Lagrangian takes the simple form:

\begin{eqnarray}
\mathcal{L} & = & -\frac{1}{2}Tr\left\{ G_{\mu\nu}G^{\mu\nu}\right\} -Tr\left\{ D_{\mu}V_{\nu}D^{\mu}V^{\nu}\right\} +Tr\left\{ D_{\mu}V_{\nu}D^{\nu}V^{\mu}\right\} \nonumber \\
 &  & -g^{2}Tr\left\{ \left[V_{\mu},V_{\nu}\right]\left[V^{\mu},V^{\nu}\right]\right\} \label{eq:LagUnit}\\
 &  & -gTr\left\{ G_{\mu\nu}\left[V^{\mu},V^{\nu}\right]\right\} +M^{2}Tr\{V_{\nu}V^{\nu}\}\nonumber \\
 &  & -\frac{1}{2\xi}\left(\partial^{\mu}G_{\mu}^{a}\right)^{2}-\frac{1}{2\xi}\left(\partial^{\mu}V_{\mu}^{a}\right)^{2}+\bar{c}^{a}\partial^{\mu}D_{\mu}^{ab}c^{b}\nonumber 
\end{eqnarray}
and the propagators are:
\begin{eqnarray}
\Delta_{G} & = & \frac{-i\delta^{ab}}{q^{2}}\left(g^{\mu\nu}+\left(\xi-1\right)\frac{q^{\mu}q^{\nu}}{q^{2}}\right)\\
\Delta_{V} & = & \frac{-i\delta^{ab}}{q^{2}-M^{2}}\left(g^{\mu\nu}+\left(\xi-1\right)\frac{q^{\mu}q^{\nu}}{q^{2}-\xi M^{2}}\right)
\end{eqnarray}

Notice that the coloron propagator is the same one we would have obtained
for a massive spin-one field in a gauge theory with spontaneous symmetry
breaking.

\subsection{Consequences of Unitarity}

Interestingly, Lagrangian (\ref{eq:LagUnit}) posses a new $Z_{2}$
symmetry under which $V$ is odd and $G$ is even. This discrete symmetry
makes the coloron to be stable. Hence, an unexpected consequence of
unitarity is that the coloron, conveniently dressed by gluons, will
form a new kind of stable hadron that can be a cold dark matter candidate.
Another implication of the $Z_{2}$ symmetry is that this kind of
coloron cannot be resonantly produced at a collider. The easiest way
to create it, is pair production. Naturally, the produced colorons
will hadronize producing two jets. Because of the huge background
for two jets at a hadron collider and the absence of any distinctive
kinematic structure, we expect that the observation of this kind of
colorons at the LHC would be very challenging.

In the ``Two-Connections'' picture, on the other hand, the $Z_{2}$
symmetry translates as a symmetry of the Lagrangian under the interchange
of the two connections ($A_{1}\leftrightarrow A_{2}$). Imposing this
symmetry to Lagrangian (\ref{eq:L2CYM1}), we obtain that the coupling
constants must satisfy the condition $g_{1}=g_{2}=\sqrt{2}g$ and
the Lagrangian (including the gauge fixing and ghost sectors) takes
the simple form:
\begin{eqnarray}
\mathcal{L} & = & -\frac{1}{2}Tr\left[F_{1\mu\nu}F_{1}^{\mu\nu}\right]-\frac{1}{2}Tr\left[F_{2\mu\nu}F_{2}^{\mu\nu}\right]+\frac{M^{2}}{2}Tr\left[\left(A_{1\mu}-A_{2\mu}\right)^{2}\right]\nonumber \\
 &  & -\frac{1}{2\xi}\left(\partial^{\mu}A_{1\mu}^{a}\right)^{2}-\frac{1}{2\xi}\left(\partial^{\mu}A_{2\mu}^{a}\right)^{2}+\frac{\bar{c}^{a}}{2}\partial^{\mu}D_{1\mu}^{ab}c^{b}+\frac{\bar{c}^{a}}{2}\partial^{\mu}D_{2\mu}^{ab}c^{b}\label{eq:2CYMUnit}
\end{eqnarray}
Obviously, Lagrangian (\ref{eq:LagUnit}) is automatically obtained
from (\ref{eq:2CYMUnit}) after the diagonalization of the mass matrix.

\section{Summary and Conclusions \label{sec:Summary-and-Conclusions}}

Finally, we have arrived to our goal and now it is time to recapitulate
our main results. First, we studied the construction of a general
local gauge theory for the coloron (with operators of dimension up
to 4). We saw that it is plagued of undetermined coupling constants,
but, nevertheless, this general theory can be a good starting point
for effective models. A direct consequence of this degree of arbitrariness
is that the expectation of the coloron interacting with gluon with
``typical QCD intensity'' is not guaranteed. In part, this theoretical
uncertainty is due to the presence of the Stueckelberg term which,
as fas as we know, has not been considered before in the context of
coloron phenomenology.

In a second step, we were able to construct a particular gauge theory
for the coloron which is BRST invariant, consistent with perturbative
unitarity. Surprisingly, this model produces propagators with acceptable
ultraviolet behavior. Additionally, the conditions imposed by unitarity
are protected by the emergence of a discrete symmetry. For all these
reasons and from the point of view of power counting, we can expect
that the theory should be renormalizable. However, a formal proof
must still be provided. 

Of course, the method described here is not the only possible construction
of a coloron model which is consistent at the quantum level. In \cite{Chivukula2011},
for example such a model is constructed by extending the QCD gauge
group to $SU(3)\times SU(3)$. In principle, one could expect the
obvious difference in the chosen gauge groups should imply different
structures at the quantum level. For example, in \cite{Chivukula2011},
it is necessary to introduce two ghost fields, one of which becomes
massive due to the symmetry breaking process, in order to fix all
the gauges present in the model. In our case, on the other hand, only
one ghost field remains because the physical coloron is not treated
as a true gauge field. Beside that, in \cite{Chivukula2011}, it is
necessary to introduce scalar fields (the would-be Goldstone bosons)
in the symmetry breaking process. Of course, in our approach, they
are completely absent. Nevertheless, those differences are only apparent
since our model can be obtained from the $SU(3)\times SU(3)$ one
if an interchange symmetry is imposed between the two groups and if
the broken symmetry is non-linearly realized in the unitary gauge
(see equation (\ref{eq:2CYMUnit})). Since the usual methodology based
on two groups is more general, the construction presented in this
paper should be viewed as a bottom-up approach which explore the minimal
conditions needed for obtaining a consistent coloron theory.

The model developed here may be phenomenologically challenging because
the observation of and stable coloron at the LHC seems to be difficult.

\section*{Acknowledgements }

In part, the present paper was motivated by very interesting questions
posed by Brenno Vallilo and Máximo Bañados during a talk given by
me at a meeting of the Chilean Physical Society. I am also very grateful
to Máximo Bañados for making me know about bigravity theories. During
the preparation of this work, I have also enjoyed very interesting
discussions with Patricio Gaete. I also want to thanks Georgios Choudalakis
for explaining me some technical aspects of jet observation at the
LHC. The present author wants to acknowledge the use of REDUCE \cite{REDUCE}
in the computation of the amplitudes cited in the text. This work
has received financial support from Fondecyt grant nº 1120346 and
Conicyt grants \textquotedblleft{}Institute for advanced studies in
Science and Technology'' ACT-119 and \textquotedblleft{}Southern
Theoretical Physics Laboratory\textquotedblright{} ACT-91. TGD.

\end{document}